\documentclass[fleqn,usenatbib]{mnras}
\usepackage{newtxtext,newtxmath}
\usepackage[T1]{fontenc}
\usepackage{graphicx}	
\usepackage{amsmath}	
\usepackage{amssymb}	

\usepackage{color}

\title[Polynomial approximation of the Lense-Thirring rigid precession frequency]{Polynomial approximation of the Lense-Thirring rigid precession frequency}

\author[De Falco \& Motta (2018)]{
Vittorio De Falco$^{1}$\thanks{E-mail: vittorio-df@issibern.ch}
\& Sara Motta$^{2}$\\
$^{1}$International Space Science Institute, Hallerstrasse 6, 3012 Bern, Switzerland\\
$^{2}$University of Oxford, Department of Physics, Astrophysics, Denys Wilkinson Building, Keble Road, Oxford OX1 3RH, UK}

\date{Accepted 8 February 2018. Received 21 December 2017; in original form 8 February 2018}
\pubyear{2017}

\begin{document}
\label{firstpage}
\pagerange{\pageref{firstpage}--\pageref{lastpage}}
\maketitle

\begin{abstract}
We propose a polynomial approximation of the global Lense-Thirring rigid precession frequency to study low frequency quasi-periodic oscillations around spinning black holes. This high-performing approximation allows to determine the expected frequencies of a precessing thick accretion disc with fixed inner radius and variable outer radius around a black hole with given mass and spin. We discuss the accuracy and the applicability regions of our polynomial approximation, showing that the computational times are reduced by a factor of $\approx70$ in the range of minutes.  
\end{abstract}


\section{Introduction}
\label{sec:introduction} 
Quasi-Periodic Oscillations (QPOs) have been discovered in the '80s in the X-ray emission from accreting stellar-mass black holes \citep[see e.g,][for a review]{Remillard06,Done07,Motta15} and weakly magnetized neutron stars \citep[see e.g.,][]{Klis89}, hosted in low mass X-ray binaries. The study of the fast time-variability through the inspection of the Fourier power density spectra reveals the presence of narrow peaks with a distinct centroid frequency, the so-called QPOs. QPOs have been often associated with accretion-related time-scales and to certain effects of strong gravity on the motion of matter around a massive object, as predicted by the theory of General Relativity \citep[see, e.g.,][for a brief review]{Motta16}. Low-frequency QPOs, with frequencies ranging from a few mHz to a few tens of Hz are commonly observed in active black hole low-mass X-ray binaries, and have been divided into three classes: Type-A, -B, and -C \citep[see e.g.,][]{Casella05}. Type-C QPOs, with frequencies in (0.1 -- 30) Hz range, are the most frequently detected \citep[][]{Belloni16}.   

While QPOs are generally thought to arise in the immediate vicinity of black holes, their physical origin remains matter of debate.  Several models have been proposed to explain the origin and the evolution of low-frequency QPOs in X-ray binaries, in particular of type-C QPOs\footnote{Several doubts still surround the origin of type-B and type-A QPOs, for which no comprehensive models has been proposed so far.}. 
Some of them invoke the motions of matter as predicted by General Relativity, and in particular the occurrence of resonances among the certain test-particle fundamental frequencies (as in the relativistic precession model, e.g., \citealt{Stella98}, \citealt{Merloni1999}) and/or among the fundamental oscillation modes of the accretion flow (as in the resonance models, see e.g., \citealt{Abramowicz2004}, \citealt{Blaes2006}, \citealt{Torok2011}), while others are based on different kinds of instabilities occurring in the accretion flow \citep[e.g.,][]{Titarchuk1999,Tagger1999,Lamb2001}\footnote{It must be noted that most of these models are purely kinematic, based on the motion of the fluid torus, and not connected to the accretion physics.}.

The relativistic precession model \citep[][]{Stella98,Stella99} was the first model to interpret type-C QPOs as the result of Lense-Thirring precession of test particles around black holes. This precession occurs as a consequence of frame-dragging, that shifts the particle orbit out of the equatorial plane. The associated frequency is referred to as \textit{test particle precession frequency}.

\cite{Ingram09} proposed a model for type-C QPOs - the 
\textit{rigid precession model} - based on similar premises. Type-C QPOs are explained as the effect of a Lense-Thirring precession of hot, geometrically thick accretion disc \citep{Abramowicz1978,Jaroszynski1980}, 
filling the region between the truncation radius of a cool, geometrically thin and optically thick Shakura and Sunyaev disc \citep{Shakura1973}. The resulting frequency is a \textit{global} frequency, which in this paper is referred to as \textit{rigid precession frequency}. 
This global rigid precession frequency corresponds to a particular type of non-axially symmetric motion of a perfect fluid around a Kerr black hole, i.e., a global epicyclic vertical $m=1$ oscillations of a fluid torus that behaves, to a first approximation, as a rigid body (see \citealt{Abramowicz2004}, \citealt{Blaes2007}, \citealt{Abramowicz2013}). 
A geometry similar to that used in the rigid precession model has been adopted also by other authors to model the Lense-Thirring precession of thick accretion flows around accreting black holes, even  though not necessarily to directly explain the origin of quasi-periodic variability  (see, among many others, e.g.,  \citealt{Fragile2007}, \citealt{Lodato2013}, \citealt{Franchini2016}).

In this paper we propose an accurate polynomial approximation to calculate the rigid precession frequencies of a thick disc with fixed inner radius and variable outer radius, centred around a black hole of given mass and spin. 
In absence of an appropriate approximation, the determination of such frequencies requires the computationally expensive calculations of two numerical integrals, which dramatically slow down any numerical code or simulation involving the aforementioned rigid precession. 

\section{Polynomial approximation method}
\label{sec:pam} 

The rigid precession model, proposed by \cite{Ingram09}, 
requires that the type-C QPO is produced via the \textit{global}, rigid precession of a toroidal structure with a fixed inner radius and a variable outer radius, which is responsible for the frequency evolution of said QPO. 
The global \textit{rigid precession frequency} of this toroidal structure can be written as:

\begin{equation} \label{EQnup}
\frac{2\pi GM}{c^3}\nu_{\rm p}(R)=\frac{\int_{R_{\rm ISCO}}^{R}\frac{r^{3-p}}{(r^{3/2}+a)^2}\left[1-\sqrt{1-\frac{4a}{r^{3/2}}+\frac{3a^2}{r^2}}\right]dr}{\int_{R_{\rm ISCO}}^{R}\frac{r^{3-p}}{(r^{3/2}+a)}dr},
\end{equation}
where $r$, the integration radius, is expressed in units of gravitational radii ($R_{\rm g}=GM/c^2$, or simply M in case G = c = 1), $M$ is the black hole mass, $a$ is the dimensionless black hole spin parameter, $p$ is an index ranging between 0 and 1, $R$ is a generic radius corresponding to the outer edge of the torus, and $R_{\rm ISCO}$ is the innermost stable circular orbit (ISCO) radius, corresponding to the inner boundary of the torus \citep[see Eq. (2.21) in][for its explicit formula]{Bardeen72}. The rigid precession frequency, $\nu_{\rm p}$, in Eq. (\ref{EQnup}) is written in dimensionless form, i.e., it is divided by the geometrical frequency $c^3/2\pi GM$.    

\begin{figure}
	\includegraphics[trim=1cm 1cm 0cm 1cm, scale=0.31]{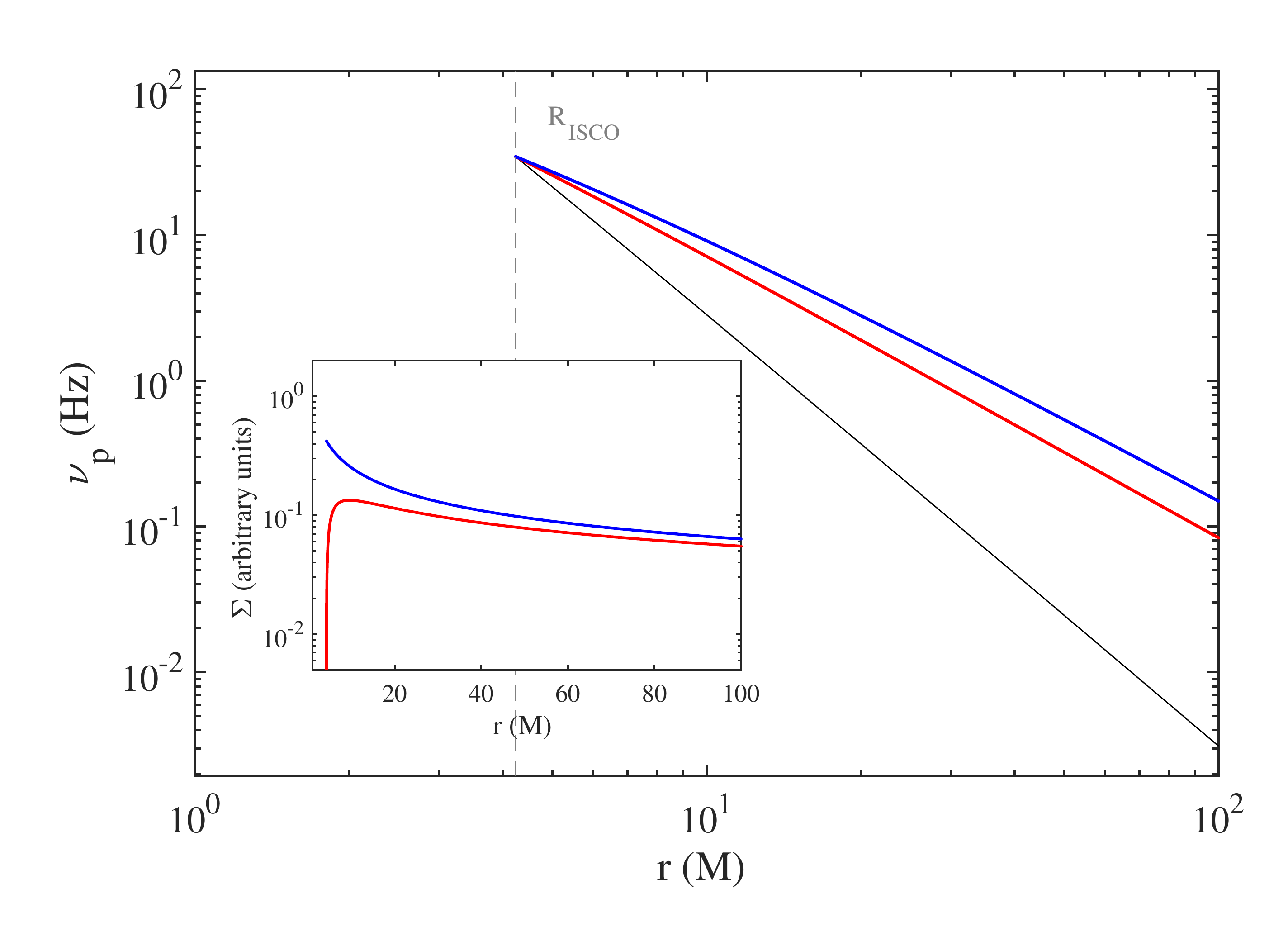}
    \caption{Lense-Thirring frequency of a rigidly precessing thick disc, for a black hole of mass $M=10 M_{\odot}$, spin $a = 0.5$, and $p = 0.6$, expressed in terms of the radius $r$. The red line represents frequencies calculated assuming a surface density profile of the form $\Sigma(r)\propto r^{-p}\left(1-\sqrt{R_{\rm ISCO}/r}\right)^p$, while the blue line is for $\Sigma(r)\propto r^{-p}$. The maximum difference in frequency is less than 2.5 Hz at about $10\,M$, and it decreases to less than 1 Hz, at $100\,M$. In the inset, the two types of density profiles are expressed in terms of the radius $r$, using the same parameter values and colours of the main figure.}
    \label{fig:Fig_sigma}
\end{figure}

In this section, we will present the procedure followed to find the polynomial approximation of the global rigid precession frequency. For this purpose, we have chosen a  simple surface density profile of the form $\Sigma(R)\propto r^{-p}$. This choice does not considerably change the computed frequencies with respect to different choices of surface density profiles \citep[see Fig. \ref{fig:Fig_sigma} and][]{Motta17}, but allows to more easily Taylor-expand, and subsequently approximate, Eq. (\ref{EQnup}) in polynomial form.

\begin{figure}
		\includegraphics[trim=0cm 0.5cm 0cm 2cm, scale=0.32]{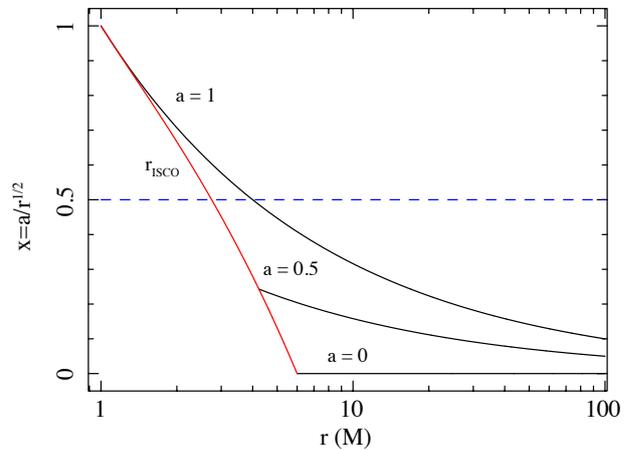}
    \caption{Parameter $x$ expressed in terms of the radius $r$, calculated for spin values of $a=0,0.5,1$. We note that in the observational spin range $a\in[0,0.5]$, $x$  (black lines) is always very small (i.e., $x<0.24$). The red line is the ISCO radius as a function of the spin. The dashed blue line, set at $x=0.5$, delimits the regions where the parameter $x$ is formally small (below the line) and high (above the line). The blue dashed line intersects the ISCO red line at $a=0.83$.}
    \label{fig:Fig1}
\end{figure}

Eq. (\ref{EQnup}) is the ratio of two elliptic integrals\footnote{An integral is defined \emph{elliptic}, when the integration term is a rational function $R\left(x,\sqrt{P(x)}\right)$ of two arguments, with $x$ being the integration variable and $P$ a polynomial of degree three or four with no repeated roots \citep[see pp. 587--607 in Chap. 17 of][for further details]{Integral64}.}, that cannot be solved analytically in terms of elementary functions. We developed a mathematical method to approximate such integrals through polynomial functions \citep[see e.g.,][]{Defalco16}. We note, that Eq. (\ref{EQnup}) can be rewritten in term of the parameter $x=a/r^{1/2}$, which will also be used to Taylor-expand such equation. This parameter can be considered small (i.e., $x\lesssim0.5$) for spins lower than $a\lesssim0.83$ (see Fig. \ref{fig:Fig1}). \cite{Motta17} found that stellar mass black holes in binary systems are generally not expected to feature spins higher than $\sim0.5$. This implies that $x$ will always range between 0 and 0.24 (see Fig.~\ref{fig:Fig1}), and can thus be considered always formally small. 
Therefore, we are allowed to expand the two integrals in Eq. (\ref{EQnup}) in Taylor series around $x=0$. By expanding the numerator up to the third order, we obtain:
\begin{equation} \label{num}
\begin{aligned}
&\int_{x_{\rm ISCO}}^{X}-\frac{2x^{2p-3}}{a^{2p-2}}\frac{1}{\left(1+\frac{x^3}{a^2}\right)^2}\left[1-\sqrt{1-\frac{4x^3}{a^2}+\frac{3x^4}{a^2}}\right]dx\approx\\
&\approx\int_{x_{\rm ISCO}}^{X}x^{2p}\left(-a^{-2p}\right)\left[4-3x-\frac{4x^3}{a^{2}}
\right]dx=\\
&=\left(-a^{-2p}\right)\left[\frac{4x^{2p+1}}{2p+1}-\frac{3x^{2p+2}}{2p+2}-\frac{4x^{2p+4}}{a^2(2p+4)}
\right]^{X}_{x_{\rm ISCO}},
\end{aligned}
\end{equation}
where $X=a/\sqrt{R}$ and $x_{\rm ISCO}=a/\sqrt{R_{\rm ISCO}}$. By expanding the denominator up to the fifth order, we obtain:
\begin{equation} \label{den}
\begin{aligned}
&\left(-\frac{2}{a^{2p-5}}\right)\int_{x_{\rm ISCO}}^{X}\frac{x^{2p-6}}{\left(1+\frac{x^3}{a^2}\right)}dx\approx\\
&\approx\left(-\frac{2}{a^{2p-5}}\right)\int_{x_{\rm ISCO}}^{X}x^{2p-6}\left[1-\frac{x^3}{a^2}
\right]dx=\\
&=\left(-\frac{2}{a^{2p-5}}\right)
\left[\frac{x^{2p-5}}{2p-5}-\frac{x^{2p-2}}{a^2(2p-2)}+\frac{x^{2p+1}}{a^4(2p+1)}
\right]^{X}_{x_{\rm ISCO}}.
\end{aligned}
\end{equation}
Therefore, the final approximation reads as:
\begin{equation} \label{Appr}
\begin{aligned}
&\frac{2\pi GM}{c^3}\nu_{\rm p}(R)\approx\frac{X^6}{2a^5}\\
&\qquad\qquad\quad\left[\frac{\frac{4}{2p+1}-\frac{3X}{2p+2}-\frac{4X^{3}}{a^2(2p+4)}
+\frac{f(x_{\rm ISCO})a^{2p}}{X^{2p+1}}}{\frac{1}{2p-5}-\frac{X^{3}}{a^2(2p-2)}+\frac{X^{6}}{a^4(2p+1)}
+\frac{g(x_{\rm ISCO})a^{2p-5}}{2X^{2p-5}}}\right],
\end{aligned}
\end{equation}
where 
\begin{eqnarray} 
f(x_{\rm ISCO})=\left(-a^{-2p}\right)
\left[\frac{4x_{\rm ISCO}^{2p+1}}{2p+1}-\frac{3x_{\rm ISCO}^{2p+2}}{2p+2}-\frac{4x_{\rm ISCO}^{2p+4}}{a^2(2p+4)}
\right],\\
g(x_{\rm ISCO})=
\left(-\frac{2}{a^{2p-5}}\right)
\left[\frac{x_{\rm ISCO}^{2p-5}}{2p-5}-\frac{x_{\rm ISCO}^{2p-2}}{a^2(2p-2)}+\frac{x_{\rm ISCO}^{2p+1}}{a^4(2p+1)}
\right],
\end{eqnarray}
with $X=a/\sqrt{R}$ and $x_{\rm ISCO}=a/\sqrt{r_{\rm ISCO}}$. We Taylor-expanded the numerator and denominator up to different orders (third and fifth, respectively) in order  to have the same number of components in both the polynomial approximations (see Eq. (\ref{Appr})), but also the minimum number of polynomial terms that allows to reach a satisfying accuracy (i.e., adding the next term in the Taylor expansion it did not show a significant improvement for all the admissible radii).   

\begin{figure}
\hbox{\includegraphics[trim=0cm 0cm 0cm 0cm, scale=0.3]{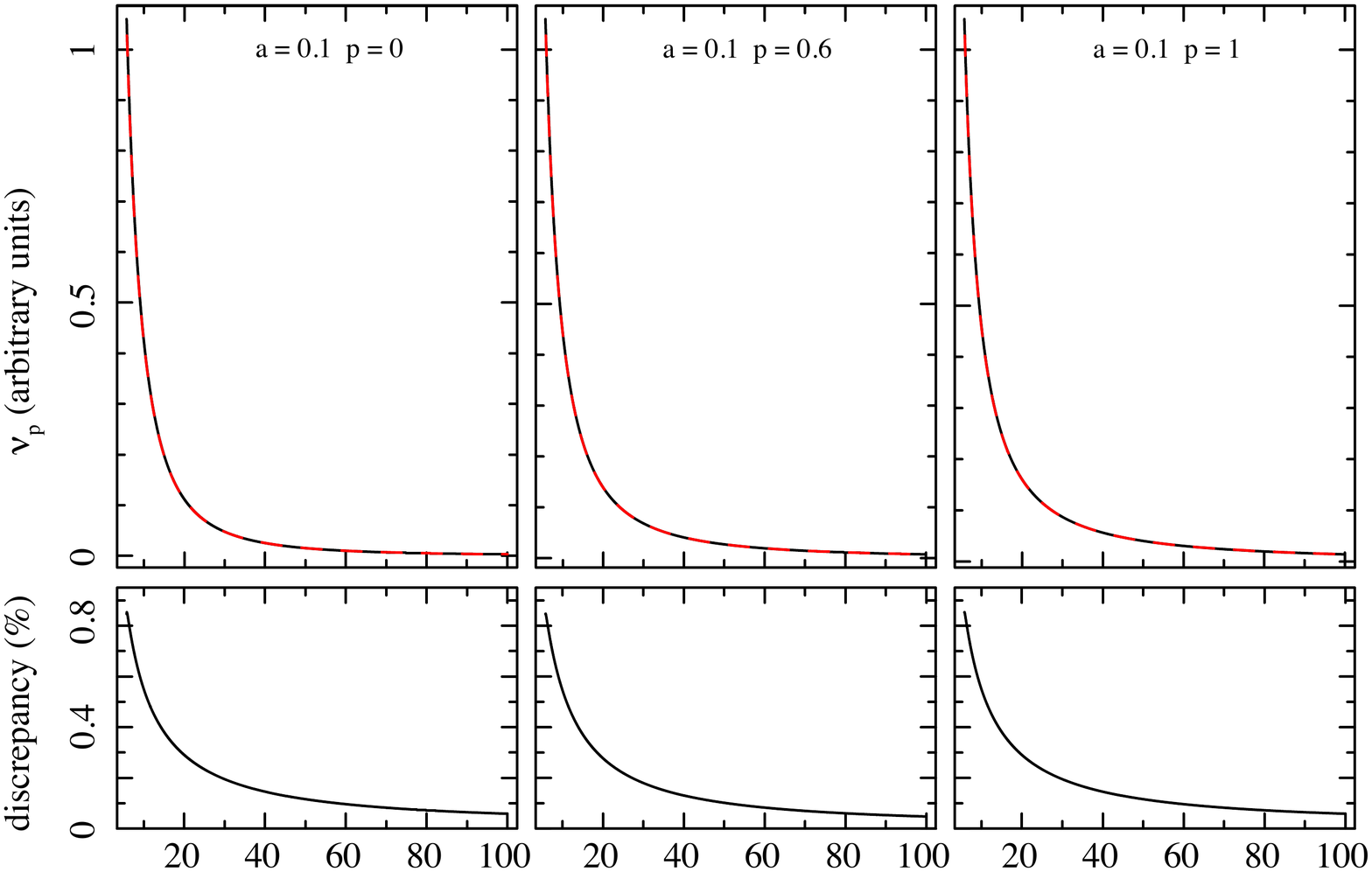}}
\vspace{-0.9cm}
\hbox{\includegraphics[trim=0cm 0cm 0cm 0cm, scale=0.3]{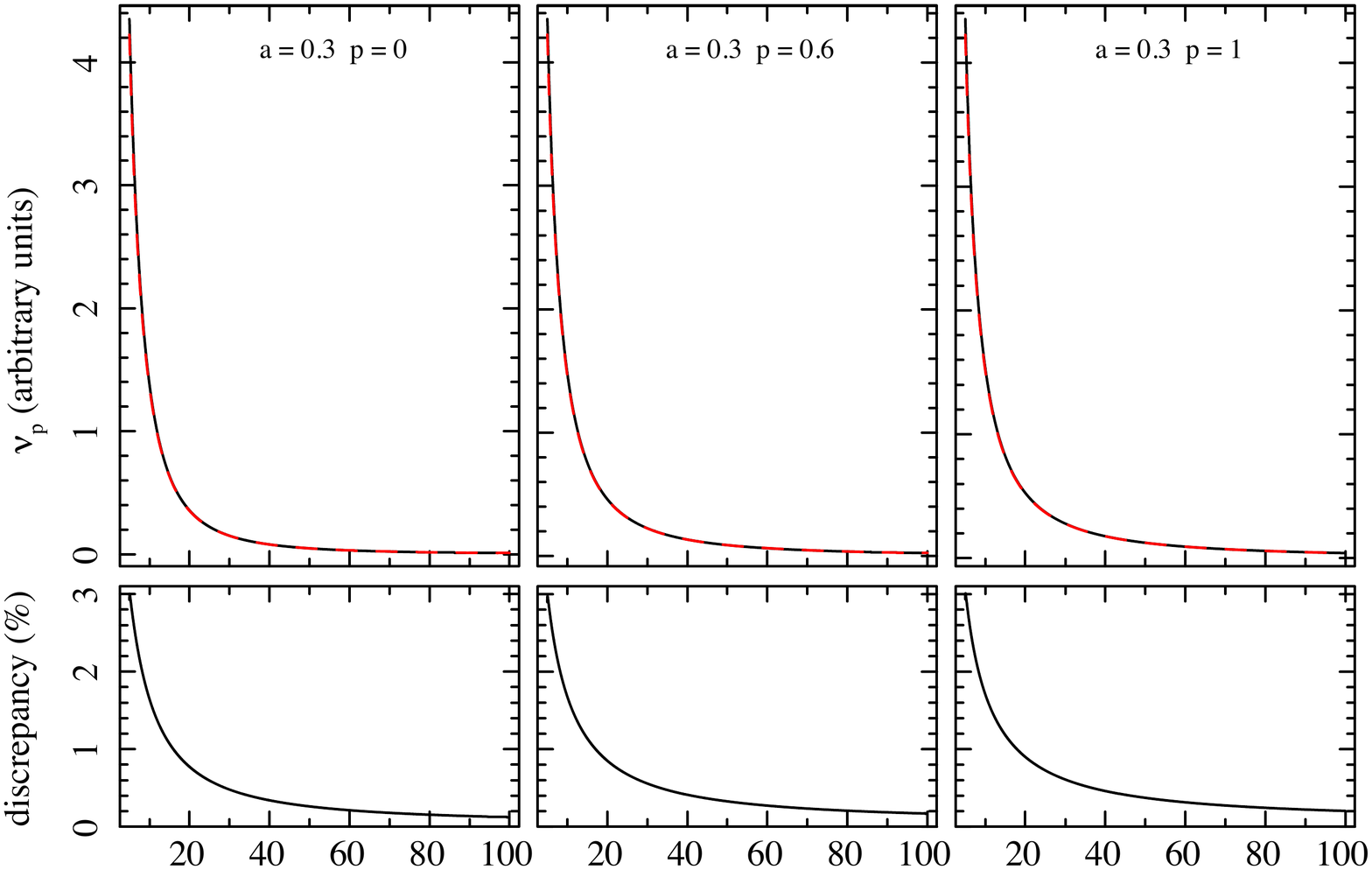}}
\vspace{-0.9cm}
\hbox{\includegraphics[trim=0cm 0cm 0cm 0cm, scale=0.3]{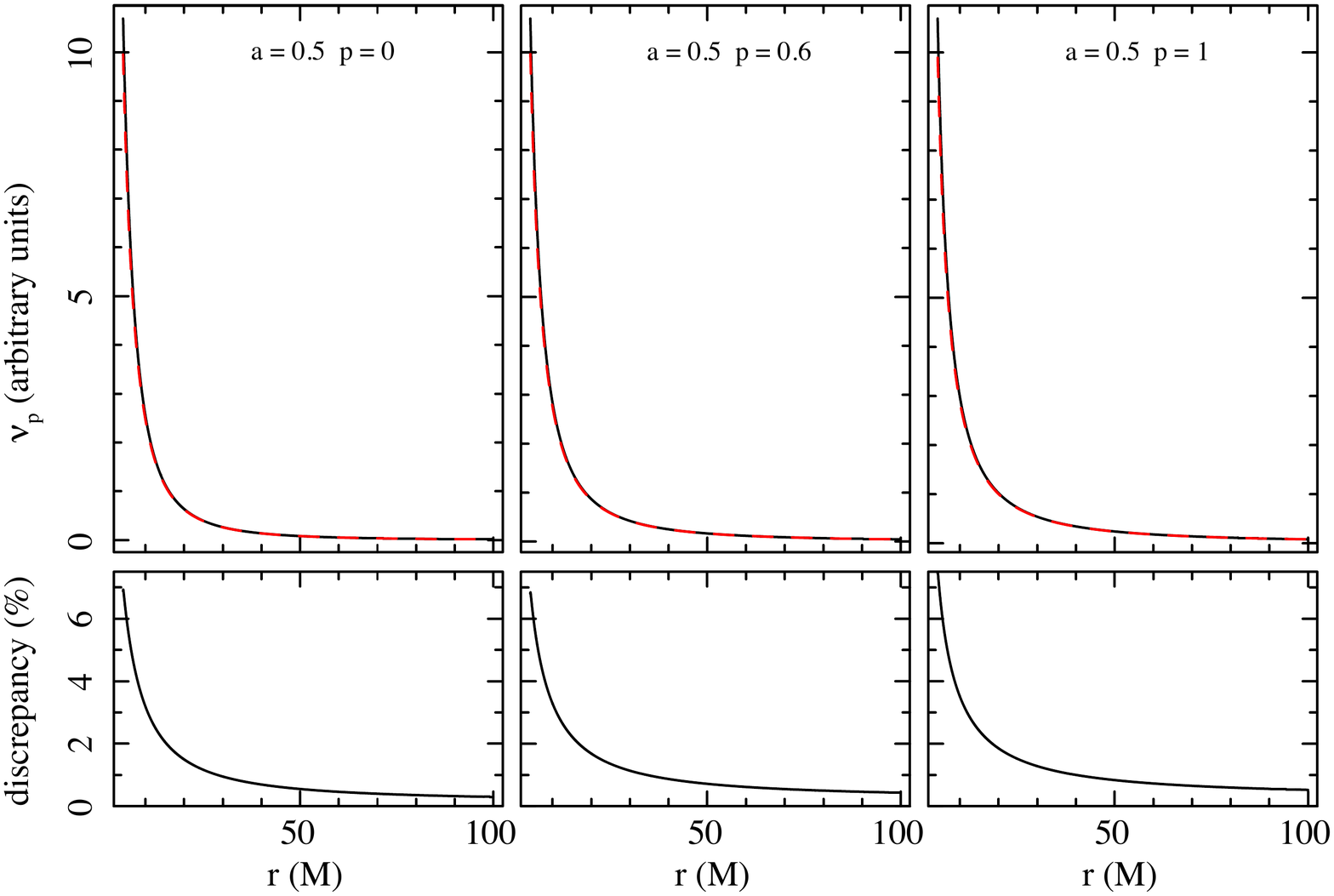}}
\vspace{-0.2cm}
\caption{Frequency expressed in arbitrary units in term of the radius, for the cases of spin values of $a=0.1,0.3,0.5$ and each one for $p=0,0.6,1$. The black continuous lines in the upper panels represent the original formula, Eq. (\ref{EQnup}), while the dashed red lines mark the approximated one, given by Eq. (\ref{Appr}). In the lower panels, we plot the discrepancy between the original and approximate formula, expressed in percentage and calculated as (|original-approximation|)/|original|, in order to estimate the accuracy of the approximation found.} 
\label{fig:Fig2} 
\end{figure}

\begin{figure}
\includegraphics[trim=0cm 0cm 0cm 0cm, scale=0.3]{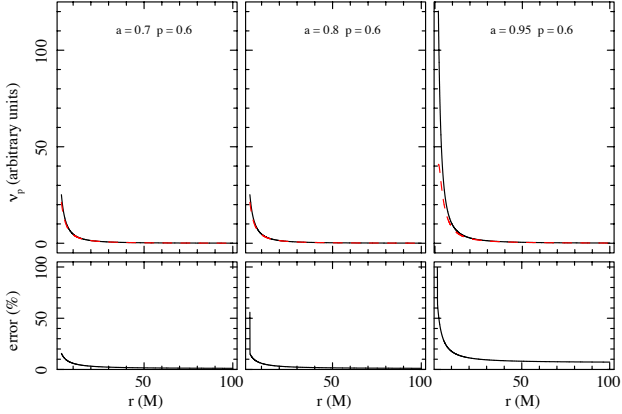}
\caption{Frequency expressed in arbitrary units in term of the radius for the cases of high spin values $a=0.7,0.8,0.95$ and $p=0.6$. The black lines in the upper panels are the original formula, Eq. (\ref{EQnup}), the red lines are the approximated one given by Eq. (\ref{Appr}). In the lower panel there is the discrepancy defined as in Fig. \ref{fig:Fig2}.} 
\label{fig:Fig3} 
\end{figure}

Fig. \ref{fig:Fig2}, displays the accuracy of the approximation found and shows that for any given set of black hole spin and mass values, the maximum discrepancy between the approximated frequency formula, Eq. (\ref{Appr}), and the original frequency formula, Eq. (\ref{EQnup}), is found at the ISCO, and decreases to zero for $R \rightarrow +\infty$. This is due to the fact that the denominator of Eq. (\ref{EQnup}) diverges to infinity for $R \rightarrow R_{\rm ISCO}$, which affects significantly the accuracy of the polynomial approximation. We note, however, that even for $a=0.5$ the maximum discrepancy, between the frequency calculated through Eq. (\ref{EQnup}) and that obtained through the approximate formula Eq. (\ref{Appr}), is of $\approx 7\%$ at $R_{\rm ISCO}$ and $a=0.5$ (and of course smaller at larger radii). Remarkably, the discrepancy profile is essentially independent from the exponent $p$, introduced via the surface density profile, but it varies with $a$, increasing steeply with it (see Fig. \ref{fig:Fig1} and Fig. \ref{fig:Fig3}). 

The asymptotic behaviour of Eq. (\ref{EQnup}) for $R\to+\infty$ (i.e., $X\to0$) can be estimated from Eq. (\ref{Appr}), that is: $\nu_{\rm p}\sim \left[\frac{(2p-5)\,f(x_{\rm ISCO})}{2a^{5-2p}}\right]\,x^{5-2p}$.
The test-particle precession frequency, $\nu_{\rm LT}$, written in terms of the parameter $X=a/\sqrt{R}$ \citep[see Eq. (1) in][]{Motta17}, for $R\to +\infty$ follows $\nu_{\rm LT}\sim (2/a^4)\,x^6$. Therefore, $\nu_{\rm p}/\nu_{\rm LT}\sim\left[\frac{(2p-5)\,f(x_{\rm ISCO})}{4a^{1-2p}}\right]\,x^{-2p-1}$. 
This asymptotic behaviour is close to $\sim R/R_{\rm ISCO}$, previously empirically estimated by \cite{Motta17}. 

\begin{figure}
\includegraphics[trim=2cm 0cm 0cm 0cm, scale=0.32]{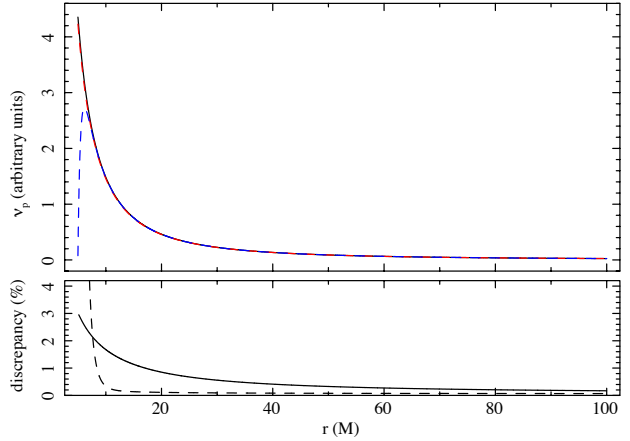}
\caption{Frequency expressed in arbitrary units in terms of the radius, for $a=0.3$ and $p=0.6$. The continuous black line in the upper panel marks the rigid precession frequency given by Eq. (\ref{EQnup}). The dashed red line is the approximated frequency, given by Eq. (\ref{Appr}), while the dashed blue line is the approximated frequency, given by Eq. (\ref{Appr2}). In the lower panel, we show the discrepancies between the original and the two approximate formulas. The continuous line corresponds to the first approximation, Eq. (\ref{Appr}), and the dashed line corresponds to the second approximation, Eq. (\ref{Appr2}). We calculated discrepancies in the same way as in Fig. \ref{fig:Fig2}.} 
\label{fig:Fig4} 
\end{figure}

Eq. (\ref{Appr}) can be further approximated and reduced to a single polynomial approximation. We define:
\begin{equation}
y=\frac{X^{3}}{a^2(2p-2)}-\frac{X^{6}}{a^4(2p+1)}
-\frac{g(x_{\rm ISCO})a^{2p-5}}{2X^{2p-5}},
\end{equation}
which is very small, being $X$ very small and $p\in[0,1]$ (which implies that $g(x_{\rm ISCO})a^{2p-5}/(2X^{2p-5})$ is also very small). By expanding $1/(1-y)$ around $y=0$ up to the third order (for the same reason explained above), we obtain:
\begin{equation} \label{Appr2}
\begin{aligned}
&\frac{2\pi GM}{c^3}\nu_{\rm p}(R)=\frac{X^6}{2a^5}\left[\frac{4}{2p+1}-\frac{3X}{2p+2}-\frac{4X^{3}}{a^2(2p+4)}\right.\\
&\left.
\qquad\qquad\qquad+\frac{f(x_{\rm ISCO})a^{2p}}{X^{2p+1}}\right](1+y+y^2+y^3+y^4).
\end{aligned}
\end{equation}
In Fig. \ref{fig:Fig4}, we compare this second approximation (given by Eq. (\ref{Appr2})) and the previous one, Eq. (\ref{Appr}). We see, that Eq. (\ref{Appr2}), blue line, provides a much less accurate approximation than Eq. (\ref{Appr}), red line, around $R_{\rm ISCO}$. This is due to the fact that we approximate twice the original formula, which induces a prominent propagation of errors (in terms of losses in accuracy), especially around $R_{\rm ISCO}$, where the denominator of the original function diverges to infinity. This second approximation behaves similarly to the first approximation with respect to parameters $a$ and $p$. This means that the discrepancies grow proportionally to $a$ and are almost independent from $p$.

\begin{figure}
\includegraphics[trim=1cm 1cm 0cm 0cm, scale=0.33]{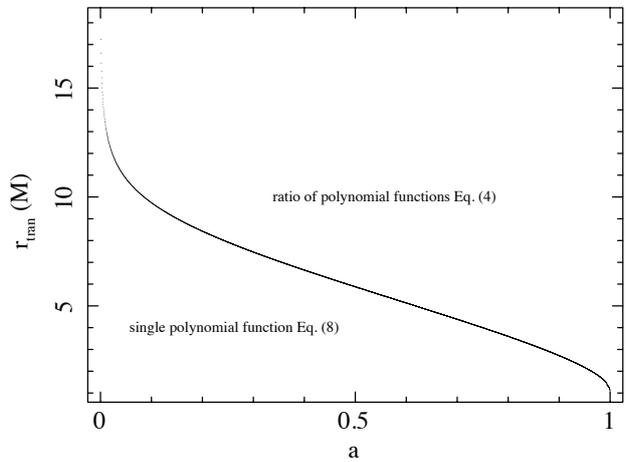}
\caption{Transition radius, $r_{\rm tran}$, expressed in terms of spin $a$. For $r<r_{\rm tran}$, Eq. (\ref{EQnup}) can be better approximated by the ratio of polynomial functions given by Eq. (\ref{Appr}), while for $r>r_{\rm tran}$ the better approximation is provided by the single polynomial function given by Eq. (\ref{Appr2}).} 
\label{fig:Fig7} 
\end{figure}

From Fig. \ref{fig:Fig4} we note that there exists a transition radius, $r_{\rm trans}$, that determines where Eq. (\ref{EQnup}) is better approximated by the ratio of polynomial functions given by Eq. (\ref{Appr}), or where it is better approximated by the single polynomial function given by Eq. (\ref{Appr2}). Such a radius is a function of $a$ and $p$, but since the computed frequencies are only weakly sensitive to $p$ (see \citealt{Motta17}), we show in Fig. \ref{fig:Fig7} only the dependence on $a$. Based on this radius, one can easily determine which approximation should be used depending on the problem considered.

\section{Concluding remarks}
\label{sec:end} 
We found an analytical approximation for the \textbf{global} rigid precession frequency of a toroidal structure centred around a spinning black hole of given mass a spin. 
In order to be able to perform the mathematical approximation, we considered a global rigid precession frequency weighted by a radial surface density profile in the form $\Sigma(r)=\Sigma_0 r^{-p}$, slightly simpler then that considered in, e.g., \cite{Motta17}. This choice is justified by the fact that such a change in the surface density profile does not strongly alter the output frequencies. 

We provided two polynomial approximations of the rigid precession frequency, in the form of ($i$) a ratio of polynomial functions, Eq. (\ref{Appr}), and ($ii$) a simple polynomial form, Eq. (\ref{Appr2}). While the former provides a better global approximation, which works well both close to ISCO and at large radii,  the latter works very at radii larger than $R_{\rm trans}$ (but not close to ISCO) and is therefore preferred when one is mostly interested in the asymptotic behaviour of the rigid precession frequency.   
The discrepancy between the original and approximate formulas (for both approximations) grows proportionally to the value of $a$, but is only weakly sensitive to the value of $p$.

The range of applicability of the approximate formulas (Eq. (\ref{Appr}) and Eq. (\ref{Appr2})) is ideally $a\in[0,0.83]$, which guarantees that the mathematical approximation we performed remains formally correct. However, as $a$ approaches its upper boundary, the analytical approximations loose accuracy. Nevertheless, we note that accreting black holes in low-mass X-ray binaries seem to have spins generally smaller than 0.5 (see discussion in \citealt{Motta17}), which implies that the approximations we found are likely suited for most realistic situations.

Our approximations enable to significantly reduce the computational times to fit the observational data and to speed-up numerical codes and theoretical calculations. As an example, we have computed the rigid precession frequencies using the original integral formula and our approximations using \textit{MATLAB} and \textit{FORTRAN 77}. In both cases, the use of the approximate formulas remarkably reduces the computational times by a factor of approximately 70 in the range of minutes. 


\section*{Acknowledgements}
VDF and SEM are grateful to the referee of this paper, Marek Abramiwicz, who contributed to improve our work with useful comments and suggestions.  VDF and SEM also thank Maurizio Falanga and Luigi Stella for the useful comments and discussions that helped to improve this paper. VDF thanks the Swiss National Science Foundation project 200021\_149865, which financed this research, the International Space Science Institute in Bern for the support, and the University of Oxford for the hospitality. SEM acknowledges the STFC for financial support.

\bibliographystyle{mnras}
\bibliography{references}
\label{lastpage}
\end{document}